\documentclass[twocolumn,prl,aps,showpacs]{revtex4}
\usepackage{graphicx}
\usepackage{xcolor}

\begin{document}

\title{Analytical Studies of the Magnetic Domain Wall Structure
in the presence of Non-uniform Exchange Bias}

\author{Yee-Mou Kao}
\email{ymkao@cc.ncue.edu.tw}
\author{Chi-Ho Cheng}
\email{phcch@cc.ncue.edu.tw}

\affiliation{Department of Physics, National Changhua University of Education, Changhua 500, Taiwan}

\date{\today}

\begin{abstract}
The pinning phenomena of the domain wall in the presence of exchange bias is studied
analytically. The analytic solution of the domain wall spin configuration is presented.
Unlike the traditional solution which is symmetric, our new solution could exhibit the
asymmetry of the domain wall spin profile. Using the solution, the domain wall position,
its width, its stability, and the depinning field are discussed analytically.
\end{abstract}

\maketitle

%----------------------------------------------------------------------
\section{Introduction}

Magnetic recording has been the most successful method for data storage in the last few decades.
In 2008, Parkin {\it et.al.} proposed a racetrack memory which has all the advantages of
magnetoresistance random access memory (MRAM) and all metallic semiconductor free structure \cite{Parkin1}. Racetrack memory consists of an ferromagnetic wire where a magnetic domain wall (DW) can be injected and detected. A 180$^{0}$ transverse DW carries a data bit via its configuration of either north
to north or south to south poles.
Several directions were also proposed to apply nanofabrication techniques to geometrically control the DW width and shape \cite{Goolaup}. Artificially induced defects could be used as pinning sites, while nanopatterned structures provide modification of the DW configuration, size and dynamical properties \cite{Parkin2}.

Recently, it was found that the pinning site, e.g., notch, may generate topological defect
and then change the chirality and topological properties of DW structure.
The chirality of DW will affect its trajectory in the Y-shape wire \cite{Parkin3,burn}.
The topological defect pinning may not be a good option for data storage.

Another option is making use of the exchange bias effect to pin the DW in
ferromagnetic material, which could be more stable and smaller in size.
As illustrated in Fig. 1, the DW is generated in ferromagnetic (F) wire \cite{nogues}.
The pinning is controlled through the exchange bias induced by the antiferromagnetic (AF) wire.
Its possibility was recently realized in experiments \cite{Polenciuc}
and simulation \cite{Albisetti}. However, its theoretical understanding is still lacking.

In extreme condition, without the magnetostatic and surface energies,
only the anisotropy and exchange energies are considered,
the spin orientation near the domain wall \cite{coey,Shibata} reads
\begin{eqnarray}
 \theta(x) = 2 \tan^{-1}\left[ \exp\left( \frac{x}{\lambda} \right) \right] \label{solu0}
\end{eqnarray}
describing a head-to-head Block wall in $x$-direction with
the spin angle $\theta(x)$ \cite{you}. $\lambda=\sqrt{A_{\rm ex}/K}$ with
the exchange stiffness $A_{\rm ex}$ and the anisotropy constant $K$ along $x$-axis.
This formula gives the domain wall width $\delta_{\rm DW} \simeq \pi \lambda$
and energy density $\varepsilon_{\rm DW} = 4 \sqrt{A_{\rm ex} K}$.

For thin magnetic nanowires, since the shape anisotropy
is mainly determined by the thickness and width of the nanowires,
the anisotropy should be perpendicular or in-plane \cite{Aharoni,DeJong}.
In this paper, only the in-plane case is considered for simplicity.
The analytic solution of the domain wall profile is obtained.
With the help of the analytic solution, the relationship between spin orientation and the
length scales of the domain wall is derived.
The position of domain wall, its width, its stability are also discussed.

%----------------------------------------------------------------------
\begin{figure}[tbh]
\includegraphics[width=3in]{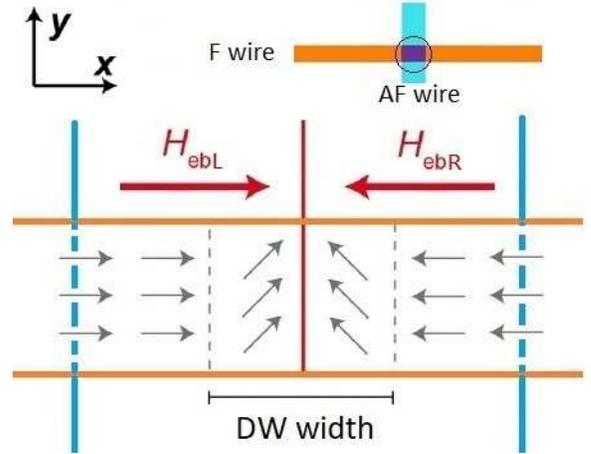}
\caption{
 Illustration of the exchange bias field (red arrows) and magnetization vectors (grey arrows)
 in F wire. The AF wire boundary is marked by blue lines.
Domain wall region is identified by grey dashed lines.
 The exchange bias field $H_{\rm ebL}$ in the left and $H_{\rm ebR}$ in the right
 are oriented at opposite directions.
}\label{eb1}
\end{figure}

%----------------------------------------------------------------------------------
{ \section{model for non-uniform exchange bias} }

{
In ferromagnetic material, two magnetic atoms interact with the so-called
exchange interaction, $J {\vec S}_1 \cdot {\vec S}_2$. $J$ is the exchange
constant, and $\vec S_1$, $\vec S_2$ are magnetic moments of two atoms.
In one dimensional wire and continuum limit, suppose the atoms only interact with
their nearest neighbors and the direction of magnetization varies slowly along the
wire, the energy, which we call the exchange energy $E_{\rm ex}$, is
\begin{eqnarray}  \label{exenergy}
E_{\rm ex} = A_{\rm ex} \int_{-\infty}^{+\infty} dx  \left(\frac{d \theta}{dx}\right)^2
\end{eqnarray}
up to a constant. $A_{\rm ex}$ which is proportional to $J$
is called the exchange constant. $\theta(x)$ is the
orientation of the magnetization at position $x$.

If we further consider the coupling between the ferromagnetic material and another antiferromagnet,
an unidirectional anisotropy would be induced in the ferromagnetic material,
which is usually referred to exchange bias \cite{Hoffmann}.
The corresponding exchange bias energy density could be modeled by
\begin{eqnarray}  \label{varepsilon}
\varepsilon_{\rm eb} = - K_{\rm eb} \cos(\theta (x) - \theta_{\rm eb})
\end{eqnarray}
where $K_{\rm eb} $ is called the unidirectional exchange coupling constant.
$\theta_{\rm eb}$ is the angle between the magnetic moment and unidirectional anisotropy axes.

In our system, as illustrated in Fig.\ref{eb1},
besides the exchange energy of F wire,
there is also the exchange bias energy $E_{\rm eb}$
due to the coupling between the F and AF wires.
At the interface between F and AF wires, the exchange anisotropy effect could create
the domain wall in F wire. As shown in Fig.\ref{eb1}, in the left (right) hand side of F wire,
the magnetization points to the right (left) due to the coupling from AF wire.
Hence
\begin{eqnarray}
  \theta_{\rm eb} =
\left\{\begin{array}{ccc}
     0 &\mbox{if} & x<0 \\
     \pi  &\mbox{if} & x>0
     \end{array} \right.
\end{eqnarray}
and $K_{\rm eb}$ are also different in the left and right sides.
If we define the exchange bias field $\vec H_{\rm eb}$ such that
its magnitude $H_{\rm eb}=K_{\rm eb}/M_{\rm s}$,
where $M_{\rm s}$ is the saturation magnetization of F wire. The direction of $\vec H_{\rm eb}$
is along the unidirectional anisotropy axes.
It follows that
\begin{eqnarray} \label{heb}
  \vec{H}_{\rm eb} =
\left\{\begin{array}{ccc}
     H_{\rm eb L} \; \hat{e}_x  &\mbox{if} & x<0 \\
    -H_{\rm eb R} \; \hat{e}_x  &\mbox{if} & x>0   \label{effecth}
      \end{array}\right.
\end{eqnarray}
where $H_{\rm eb L}$ and $H_{\rm eb R}$ are the exchange bias field intensities
in the left and right regions, respectively.
Domain wall width from 150 nm to 1 $\mu$m range
can be obtained at the boundary between two regions with opposite exchange bias field ranging from 50 to 300 Oe. These exchange bias values are compatible with those found in the
Fe$_{40}$Co$_{40}$B$_{20}$/Ir$_{20}$Mn$_{80}$ or Py/Ir$_{20}$Mn$_{80}$ systems \cite{YDu}.

Eq.(\ref{varepsilon}) can then be re-written as
\begin{eqnarray}
  \varepsilon_{\rm eb} = - \vec{M} \cdot \vec{H}_{\rm eb}
\end{eqnarray}
where ${\vec M}= M_{\rm s}({\hat e}_x \cos\theta(x) + {\hat e}_y \sin\theta(x))$.
It turns out that the exchange bias energy
\begin{eqnarray}  \label{exbiasenergy}
E_{\rm eb} = -\int_{-\infty}^{+\infty} dx \vec{M} \cdot \vec{H}_{\rm eb}
\end{eqnarray}

The pinning DW by exchange bias with two regions characterized by different unidirectional anisotropy
was proposed by Albisetti {\it et.al.} \cite{Albisetti}

}

%----------------------------------------------------------------------
{ \section{Domain Wall Structure} }

{
Combining the exchange energy $E_{\rm ex}$ in Eq.(\ref{exenergy}) and
the exchange bias energy $E_{\rm eb}$ in Eq.(\ref{exbiasenergy}), we get
the DW energy }
\begin{eqnarray}
E =  A_{\rm ex} \int_{-\infty}^{+\infty} dx  \left( \frac{d \theta}{dx} \right)^2
-\int_{-\infty}^{+\infty} dx \vec{M} \cdot \vec{H}_{\rm eb}
\end{eqnarray}
The DW profile is determined by their competition.
Decompose the DW energy into two regions,
\begin{eqnarray}
 E &=& \int_{0}^{\infty} dx \left[ A_{\rm ex}\left( \frac{d \theta}{d x} \right)^2
       + M_{\rm s} H_{\rm ebR} \cos\theta \right]  \nonumber \\
 && +\int_{-\infty}^{0} dx \left[ A_{\rm ex} \left( \frac{d \theta}{d  x} \right)^2
      - M_{\rm s} H_{\rm ebL} \cos\theta \right]
     \label{totalenergy}
\end{eqnarray}
Minimization with respect to $\theta(x)$ gives
\begin{eqnarray}
2A_{\rm ex}\frac{d^2 \theta}{d x^2} -M_{\rm s} H_{\rm ebL}\sin\theta = 0 \;\;\;\mbox{if}\;\;\; x<0   \\
2A_{\rm ex}\frac{d^2 \theta}{d x^2} +M_{\rm s} H_{\rm ebR} \sin\theta = 0 \;\;\;\mbox{if}\;\;\; x>0  \label{eulerdiff}
\end{eqnarray}
with the boundary conditions
\begin{eqnarray} \label{bc}
\lim_{x\rightarrow -\infty}\theta(x) &=& 0  \\
\lim_{x\rightarrow +\infty}\theta(x) &=& \pi
\end{eqnarray}
and further the continuity imposed at $x=0$,
says, $\theta(x=0) = \theta_0$ as an undetermined parameter.
The solution is found to be
\begin{widetext}
\begin{eqnarray} \label{solu}
\theta(x) = \left\{\begin{array}{lcc}
4 \tan^{-1}\left[\tan\left( \frac{\theta_0}{4}\right)
         \exp\left(\frac{x}{\lambda_{\rm L}}  \right)
	     \right]  &\;\;\mbox{if}\;\;& x<0  	     \\
\pi - 4 \tan^{-1}\left[ \tan\left( \frac{\pi-\theta_0}{4}\right)
         \exp\left(-\frac{x}{\lambda_{\rm R}}   \right)
         \right]  &\;\;\mbox{if}\;\;& x>0
         \end{array}\right. \label{solu1}
\end{eqnarray}
\end{widetext}
where $\lambda_{\rm L}=\sqrt{2A_{\rm ex} /(M_{\rm s} H_{\rm eb L})}$ and
$\lambda_{\rm R}=\sqrt{2A_{\rm ex} /(M_{\rm s} H_{\rm eb R})}$ define the length scales
of the domain wall in the left and right regions, respectively.
Here we obtained a formula different from the traditional one used in micromagnetics
as shown in Eq.(\ref{solu0}). The traditional formula is applied for head-to-head Block wall
whereas it is the Neel wall in our case. The spin orientation at $x=0$, $\theta_0$,
is determined by the continuity of its derivatives,
{\rm i.e.}, $\theta'(x=0^-)=\theta'(x=0^+)$, which gives
\begin{eqnarray}
 \theta_0  = 2 \tan^{-1}\left( \frac{ \lambda_{\rm L}}{\lambda_{\rm R}} \right)
 \label{theta0}
\end{eqnarray}
If the bias field is symmetric, {\rm i.e.}, $H_{\rm ebL}=H_{\rm ebR}$,
then $\lambda_{\rm L} = \lambda_{\rm R}$, $\theta_0 = \pi/2$, and
obviously the DW center $x_{\rm c}=0$ by symmetry. In general, the bias field is
not neccssary to be symmetric, {\rm i.e.}, $H_{\rm ebL}\neq H_{\rm ebR}$,
the DW width becomes $\delta_{\rm DW} = \pi(\lambda_{\rm L} + \lambda_{\rm R})/2$ \cite{dwwidth}.
The DW center, $x_{\rm c}$, defined as the position such that $\theta(x_{\rm c})=\pi/2$,
can be found by using Eqs.(\ref{solu1})-(\ref{theta0}), which gives
\begin{widetext}
\begin{eqnarray}
  x_{\rm c} =
\left\{\begin{array}{lcc}
     \lambda_{\rm R} \left[ \ln\left( \sqrt{2} + 1 \right)
     - \ln\left( \frac{\sqrt{\lambda_{\rm L}^2+\lambda_{\rm R}^2}
     + \lambda_{\rm L}}{\lambda_{\rm R}} \right) \right] &\mbox{if}& \;  H_{\rm ebL} > H_{\rm ebR} \\
     -\lambda_{\rm L} \left[ \ln\left( \sqrt{2} + 1 \right)
     - \ln\left( \frac{\sqrt{\lambda_{\rm L}^2+\lambda_{\rm R}^2}
     + \lambda_{\rm R}}{\lambda_{\rm L}} \right) \right] &\mbox{if}& \;  H_{\rm ebL} < H_{\rm ebR}
      \end{array}\right.   \label{xc}
\end{eqnarray}
\end{widetext}

If the lowest order is kept, the expression can be simplified as
\begin{eqnarray}
x_{\rm c} = \frac{1}{\sqrt{2}} ( \lambda_{\rm R} - \lambda_{\rm L} )
\end{eqnarray}
for $|\lambda_{\rm L}-\lambda_{\rm R}|\ll \lambda_{\rm L}$ and
$|\lambda_{\rm L}-\lambda_{\rm R}|\ll \lambda_{\rm R}$.
This serves as a useful formula for fast estimation of the domain wall center position.
The spin orientation $\theta(x)$ along the F wire for different bias is shown in Fig.2.
It can be seen that as bias field asymmetry increases, the domain wall becomes wider, and the
domain wall center will shift to the direction of lower bias.
{ It implies that one can
fine-tune the DW position and modifies the DW width through exchange bias.}
To quantify their changes,
let the dimensional parameter $h=\frac{H_{\rm ebL}-H_{\rm ebR}}{H_{\rm ebL}+H_{\rm ebR}}$
to represent the degree of asymmetry bias.
The DW width and the center position can then be re-written as
\begin{eqnarray}
 \delta_{\rm DW}
 &=& \frac{\pi}{2} \sqrt{\frac{A_{\rm ex}}{M_{\rm s} ( H_{\rm ebL}+H_{\rm ebR} )}}
  \left[  \left( 1+ h  \right)^{-1/2}  + \left( 1 - h  \right)^{-1/2} \right]  \nonumber \\
&=&  \pi \sqrt{\frac{A_{\rm ex}}{M_{\rm s} ( H_{\rm ebL}+H_{\rm ebR} )}}
 \left[1  + \frac{3}{8} h^2+ \frac{35}{128} h^4    + O\left( h^6 \right) \right] \nonumber \\
\end{eqnarray}
and
\begin{widetext}
\begin{eqnarray}
x_{\rm c} =  \sqrt{\frac{A_{\rm ex}}{2 M_{\rm s} ( H_{\rm ebL}+H_{\rm ebR} )}}
h \left( 1 + \frac{1}{4} |h| +\frac{13}{24} |h|^2 + O( |h|^3) \right)
\end{eqnarray}
\end{widetext}
The plots of their relations with $h$ are shown in Fig.3.

If an external magnetic field $\vec{H}_{\rm ext} = H_{\rm ext} \hat{e}_x$
is applied along the F wire, the bias asymmetry is modified, and so
{ it turns
out to be described by an effective exchange bias field ${\vec H}_{\rm eb}^{\rm eff}$
which is the sum of $\vec H_{\rm eb}$ from Eq.(\ref{heb}) and $\vec H_{\rm ext}$, {\rm i.e.}, }
\begin{eqnarray}
  \vec{H}_{\rm eb}^{\rm eff} =
\left\{\begin{array}{ccc}
     \left( H_{\rm eb L} + H_{\rm ext} \right)  \; \hat{e}_x  & \mbox{if} & x<0 \\
     \left( -H_{\rm eb R} + H_{\rm ext}  \right) \; \hat{e}_x & \mbox{if} & x>0   \label{effecth2}
      \end{array}\right.
\end{eqnarray}
When the applied field $H_{\rm ext}$ approaches to exchange bias in the right region,
$H_{\rm ebR}$,
the corresponding DW width in the right, which is described by the length scale
 $\sqrt{2A_{\rm ex}/(M_{\rm s}(H_{\rm ebR}-H_{\rm ext}))}$, will diverge.
Physically it implies the domain wall becomes unstable.
{ Such a critical external field
\begin{eqnarray}
 H_{\rm c} = H_{\rm ebR} \label{dep1}
\end{eqnarray}
should correspond to the depinning field with the same order of magnitude. }
It is consistent with the experimental observation that the wider AF wires, the larger
exchange bias, and hence the larger depinning field \cite{Polenciuc}.

The variation of $E_{\rm eb}$ is justified in polycrystalline exchange bias
systems characterized by large antiferromagnetic uniaxial anisotropy \cite{Grady}.
In Fe$_{40}$Co$_{40}$B$_{20}$/Ir$_{20}$Mn$_{80}$ system, the typical values of
saturation magnetization  $M_{\rm s} = 750$ kA/m, exchange stiffness $A_{\rm ex} = 1.2 \times 10^{-11}$J/m,
If $H_{\rm eb}$ is 175 Oe, then $\lambda_H \simeq 42.8$ nm.
The unidirectional anisotropy constant $K_{\rm eb}$ = 6.56 kJ/m$^3$.
then the domain width $\delta_{\rm DW} \simeq \pi\lambda_{H}$ = 134 nm.
The energy density is $\gamma_{\rm DW} \simeq 4\sqrt{A_{\rm ex} K_{\rm eb}/2}  \simeq$ 1.12 mJ/m$^2$.

Except the exchange energy and the exchange bias energy, there are other types
of interaction involved in reality, for example, the dipolar interaction
which is of at least one order lower \cite{Kim}.
The shape anisotropy constant  $K_{\rm sh} = \mu_0 M_S^2/2$ = 0.35 MJ/m$^3$ due to
demagnetizing energy is much larger than the unidirectional anisotropy $E_{\rm eb}$ = 6.56 kJ/m$^3$
 due to exchange bias energy.
  Although $K_{\rm sh}$ much large $K_{\rm eb}$, in the nano thin, narrow strips,
 the strong demagnetizing field force the magnetization vector parallel to the plane of thin, narrow strips,
 so that the exchange bias acts as a slight modulation.
This peculiar asymmetric configuration can be obtained experimentally by ion irradiation
techniques, by modulating the ions dose for selectively destroying
or weaken the exchange coupling between the antiferromagnetic
and ferromagnetic layers and therefore the exchange bias has asymmetry \cite{Mouqin,albisetti-nt}.

%----------------------------------------------------------------------
\begin{figure}[tbh]
\includegraphics[width=3.5in]{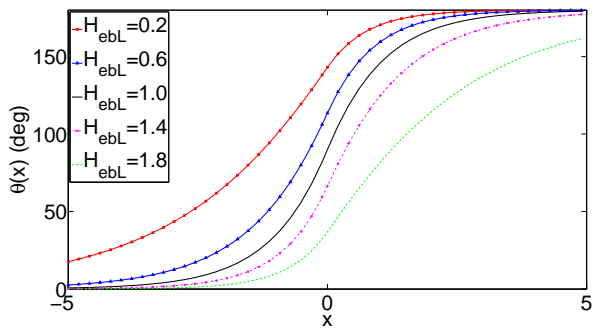}
\caption{Spin orientation $\theta(x)$ as a function of $x$ in the unit of length scale
$\sqrt{\frac{A_{\rm ex}}{M_{\rm s} ( H_{\rm ebL}+H_{\rm ebR} )}}$
for different $H_{\rm ebL}$ in the unit of $( H_{\rm ebL} + H_{\rm ebR})/2$.
 }\label{dw_asymmetry}
\end{figure}
%----------------------------------------------------------------------

%----------------------------------------------------------------------
\begin{figure}[tbh]
\includegraphics[width=3.5in]{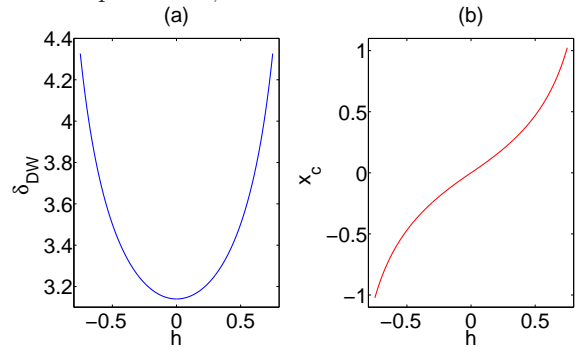}
\caption{
(a) Domain wall width $\delta_{\rm DW}$ and
(b) center position $x_{\rm c}$, in the unit of length scale
$\sqrt{\frac{A_{\rm ex}}{M_{\rm s} ( H_{\rm ebL}+H_{\rm ebR} )}}$,
as a function of $h$.
}\label{dw_bias}
\end{figure}

%----------------------------------------------------------------------
\section{Uniaxial Anisotropy}

In this section, we study the effect of in-plane uniaxial anisotropy on the DW structure,
its stability and the depinning field in the one-dimensional wire in the presence of exchange bias.

The in-plane anisotropy should play an important role in determining the domain wall structure
and also its width. In particular, the domain wall width decreases (increases) if the anisotropy
is parallel (perpendicular) to the easy axis \cite{porter,Bryan,hertel}.

{
Let $\hat{n}_a =  {\hat e}_x \cos\theta_a + {\hat e}_y \sin\theta_a $
be the direction of the easy axis due to uniaxial anisotropy,
the magnetization will prefer both $\theta_a$ and also its reverse direction $\pi-\theta_a$,
the anisotropy energy up to the leading order \cite{coey} could be represented by
}
\begin{eqnarray}
E_{\rm ani} =  - K_{\rm ani} \int_{-\infty}^{\infty}dx \cos^2(\theta-\theta_a) \label{energyani}
\end{eqnarray}
where $K_{\rm ani} > 0$ is the uniaxial anisotropy constant \cite{Grady}.
Similarly, the spin orientation is obtained by minimizing the total energy, which turns out to be
\begin{widetext}
\begin{eqnarray}
2A_{\rm ex}\frac{d^2 \theta}{d  x^2} - M_{\rm s} H_{\rm ebL}\sin\theta - K_{\rm ani} \sin 2(\theta-\theta_a)
 = 0 && \;\;\mbox{if}\;\; x<0  \\
2A_{\rm ex}\frac{d^2 \theta}{d  x^2} + M_{\rm s} H_{\rm ebR}\sin\theta - K_{\rm ani} \sin 2(\theta-\theta_a)
 = 0 && \;\;\mbox{if}\;\; x>0
\end{eqnarray}
\end{widetext}
In the following, the symmetric bias ($H_{\rm ebL}=H_{\rm ebR}=H_{\rm eb}$) is assumed
in order to understand the anisotropic effect.
Since no closed form solution of the above differential equation is found,
we adopt the solution form in Eq.(\ref{solu1}) for the case that the anisotropy energy
is small compared with the exchange bias energy, {\rm i.e.},
$K_{\rm ani} \ll M_{\rm s} H_{\rm eb}$. The domain wall length scale
$\lambda_{\rm L}=\lambda_{\rm R}=\lambda$ is left as the variational parameter.
The total energy becomes
\begin{widetext}
\begin{eqnarray}
E &=& A_{\rm ex} \int_{-\infty}^{+\infty} dx  \left( \frac{d \theta}{dx} \right)^2
-\int_{-\infty}^{+\infty} dx \vec{H}_{\rm eb} \cdot \vec{M}
+ K_{\rm ani}\int_{-\infty}^{\infty}dx \cos^2(\theta-\theta_a) \nonumber \\
&=&  4(2-\sqrt{2}) \frac{A_{\rm ex}}{\lambda}
       +  2(2-\sqrt{2})  M_{\rm s} H_{\rm eb} \lambda
       +  \frac{4}{3}(4-\sqrt{2}) K_{\rm ani} \lambda  \cos 2\theta_a
\end{eqnarray}
\end{widetext}
Minimization with respect to $\lambda$ gives
\begin{eqnarray} \label{lambda}
\lambda = \sqrt{\frac{6 A_{\rm ex}}{ 3 M_{\rm s} H_{\rm eb}
+ 2(3+\sqrt{2})  K_{\rm ani} \cos 2\theta_a  }}  \label{lambdak}
\end{eqnarray}
To compare with the simulation result \cite{Albisetti}, we set the same values
of $K_{\rm ani}$ as used in simulation. The spin orientation $\theta(x)$
for different $H_{\rm eb} / K_{\rm ani}$ is shown in Fig.4.
{
It shows that the larger anisotropy, the larger DW width.
The domain wall length scale $\lambda$ (same order of magnitude as domain wall width)
as a function of $H_{\rm eb}$ for different anisotropies is shown in Fig.\ref{eb2}.
It can be seen that the difference in DW width for different anisotropies
is insignificant if the exchange bias is large enough.
It implies that for large exchange bias, the structure of DW would be
slightly modified by anisotropy effect.}
Our result is consistent with simulation from which the same plot is shown in Fig.4(a)
in Ref.\cite{Albisetti}.

If the anisotropy effect takes place along $y$-axis,
once $K_{\rm ani} \gtrsim M_{\rm s} H_{\rm eb}$,
the domain wall width is sufficiently large such that the boundary conidtion
imposed in Eq.(\ref{bc}) becomes invalid.

{
If the external magnetic field ${\vec H}_{\rm ext}=H_{\rm ext}\hat{e}_x$ is applied,
similar to the case in previous section, we could replace $H_{\rm eb}$ by
the effective one, {\rm i.e.}, $H_{\rm eb}^{\rm eff}=-H_{\rm eb}+H_{\rm ext}$  in the right side.
The solution in Eq.(\ref{solu1}) becomes physically unstable when
the DW length scale, $\lambda$, in Eq.(\ref{lambda}), diverges.
At this moment, $H_{\rm eb}^{\rm eff}=-H_{\rm c}+H_{\rm ext}$.
It defines the critical field
\begin{eqnarray}
H_{\rm c} = H_{\rm eb} + \frac{2(3+\sqrt{2})}{3} \frac{K_{\rm ani}}{M_{\rm s}}
\cos 2\theta_a
\end{eqnarray}
which should correspond to the depinning field with the same order of magnitude.
}

%----------------------------------------------------------------------
\begin{figure}[tbh]
\includegraphics[width=3.5in]{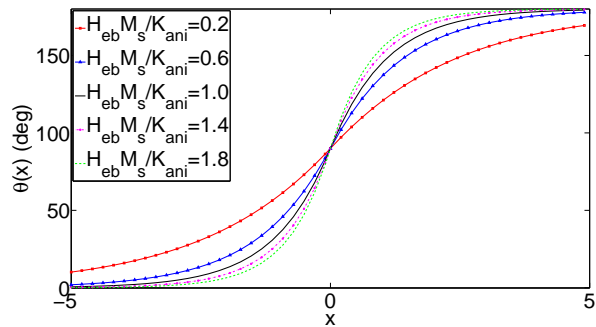}
\caption{
Spin orientation $\theta(x)$ as a function of $x$, in the unit of
$\sqrt{\frac{A_{\rm ex}}{K_{\rm ani}}}$,
for different $\frac{H_{\rm eb}M_{\rm s}}{K_{\rm ani}}$.
$K_{\rm ani} = 10^3$J/m$^3$ (along $x$-axis).
}\label{aniso_strength}
\end{figure}
%----------------------------------------------------------------------

%----------------------------------------------------------------------
\begin{figure}[tbh]
\includegraphics[width=3.5in]{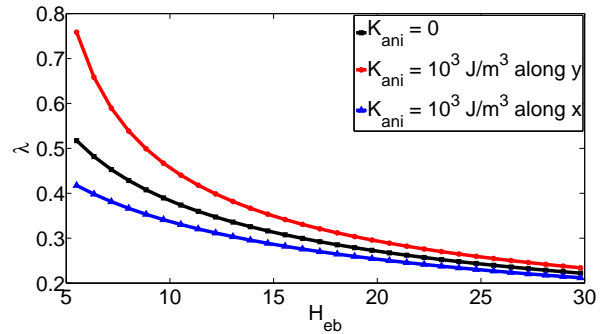}
\caption{
Domain wall length scale, $\lambda$, in the unit of
$\sqrt{\frac{A_{\rm ex}}{K_{\rm ani}}}$, as a function of exchange bias field $H_{\rm eb}$
for $K_{\rm ani}= 0$, $10^3$J/m$^3$ (along $x$-axis, $\theta_a=0$), and
$10^3$J/m$^3$  (along $y$-axis, $\theta_a=\pi/2$).
}\label{eb2}
\end{figure}
%----------------------------------------------------------------------

%----------------------------------------------------------------------
\section{Conclusion}

 We solve analytically for the spin orientation along the wire
 in the presence of non-uniform exchange bias \cite{Albisetti}, as shown in Eq.(\ref{solu}).
 Even for symmetry exchange bias field, the solution we get is different from the traditional one, as shown in Eq.(\ref{solu0}), usually appeared in the field of micromagnetics \cite{coey}.

 For asymmetry exchange bias field, the spin orientation $\theta_0$, and the center position of
 the domain wall $x_{\rm c}$ as a function domain wall length scales $\lambda_{\rm R}$ and $\lambda_{\rm R}$
 are also derived analytically. These variables can be easily measured in experiments and hence it could be verified in practise.
 Finally, with small anisotropic effect, the domain wall stability condition
 and  the depinning field are also obtained.

Although the model is so simplified that
only the exchange bias, the exchange energy, and the anisotropy effect are considered,
and the other contribution from dipolar interaction, the imperfect and edges energy,
which are at least one order lower \cite{Kim}, are ignored,
our analytic results are still consistent with previous simulation \cite{Albisetti}.
The creation and fine tuning of the domain wall by exchange bias and uniaxial anisoropy
are shown to be possible.
These results should be helpful for the development of new DW-based magnetic devices and architectures.

\section{acknowledgments}
The authors thank Lance Horng and Deng-Shiang Shiung for discussion.
The work was supported by the Ministry of Science and Technology of the
Republic of China.

\section{Data Availability Statement}
Data sharing is not applicable to this article as no new data were created or analyzed in this study.

%----------------------------------------------------------------------
%\begin{figure}[hbt]
%\includegraphics[width=3.5in]{Comparison_two_spin_angle.eps}
%\caption{
% Compare the transitional domain wall shape eq.(\ref{solu0})
% with the new domain wall shape eq.(\ref{solua},\ref{solub}) in the same parameter $\lambda=\lambda_{\rm %H}$.
% These two figures are similar, the latter has smaller width of domain wall $\delta_W = \pi \lambda$.
% }\label{dw_comparison}
%\end{figure}

%----------------------------------------------------------------------


\begin{references}
\bibitem{dwwidth}
Strictly speaking, $\pi(\lambda_{\rm L}+\lambda_{\rm R})/2$ is not exactly the domain wall width
because the magnetization direction only asymptotically approaches $0$ or $\pi$.
However, we could still define the width as its corresponding length scale
$(\lambda_{\rm L}+\lambda_{\rm R})/2$ multiplied by a constant $\pi$.
This proportionality constant is adopted to be consistent with the usual definition.
See Eq.(7.16) in Ref.\cite{coey}.
\end{references}
\end{document}